\documentstyle[aps,epsf,tighten,epsfig,amsmath,multicol]{revtex}
\renewcommand{\narrowtext}{\noindent\begin{multicols}{2}\noindent
\global\columnwidth20.5pc}
\renewcommand{\widetext}{\end{multicols}
\global\columnwidth42.5pc}  
\def\bi{{\bf i}}
\def\bj{{\bf j}}
\def\bk{{\bf k}}

\def\bq{{\bf q}}
\def\bQ{{\bf Q}}

\def\b0{{\bf 0}}

\def\up{\uparrow}
\def\down{\downarrow}

\def\eps{\epsilon}

\def\Gam{\Gamma}

\def\Lam{\Lambda}

\def\sg{\sigma}

\begin{document}
\title{$d$-wave superconductivity and Pomeranchuk instability \\ 
       in the two-dimensional Hubbard model}
\author{Christoph J. Halboth and Walter Metzner \\
{\em Institut f\"ur Theoretische Physik C, Technische Hochschule Aachen} \\
{\em Templergraben 55, D-52056 Aachen, Germany}}
\date{\small\today}
\maketitle
\begin{abstract}
We present a systematic stability analysis 
for the two-dimensional Hubbard model, which is based on a new
renormalization group method for interacting Fermi systems.
The flow of effective interactions and susceptibilities confirms
the expected existence of a $d$-wave pairing instability driven by 
antiferromagnetic spin fluctuations.
More unexpectedly, we find that strong forward scattering interactions
develop which may lead to a Pomeranchuk instability breaking the
tetragonal symmetry of the Fermi surface. \\
\noindent
\mbox{PACS: 71.10.Fd, 71.10.-w, 74.20.Mn} \\
\end{abstract}

\narrowtext


The two-dimensional Hubbard model \cite{Mon} has attracted much
interest as a promising prototype model for the electronic degrees 
of freedom in the copper-oxide planes of high-temperature 
superconductors, since it has an anti\-ferromagnetically ordered 
ground state at half-filling and is expected to become a $d$-wave
superconductor for slightly smaller electron concentrations 
\cite{Sca}.

Although the Coulomb interaction in the cuprate superconductors
is rather strong, the tendency towards antiferromagnetism
and $d$-wave pairing is captured already by the 2D Hubbard model 
at {\em weak}\/ coupling. 
Conventional perturbation theory breaks down for densities
close to half-filling, where competing infrared divergences
appear as a consequence of Fermi surface nesting and van Hove
singularities \cite{Sch1,Dzy1,LMP}.
A controlled and unbiased treatment of these divergencies cannot
be achieved by standard resummations of Feynman diagrams, but
requires a {\em renormalization group} (RG) analysis which takes 
into account the particle-particle and particle-hole channels on 
an equal footing.

Early RG studies of the two-dimensional Hubbard model started with 
simple but ingenious scaling approaches, very shortly after the 
discovery of high-$T_c$ superconductivity \cite{Sch1,Dzy1,LMP}. 
These studies focussed on dominant scattering processes between 
van Hove points in k-space, for which a small number of running 
couplings could be defined and computed on 1-loop level. 
Spin-density and superconducting instabilities where identified
from divergencies of the corresponding correlation functions.

A major complication in two-dimensional systems compared to one
dimension is that the effective interactions cannot be parameterized
accurately by a small number of running couplings, even if 
irrelevant momentum and energy dependences are neglected, since 
the {\em tangential}\/ momentum dependence of effective interactions 
along the Fermi surface is strong and important in the low-energy 
limit.
This has been demonstrated in particular in a 1-loop RG
study for a model system with two parallel flat Fermi surface 
pieces \cite{ZYD}.
Zanchi and Schulz \cite{ZS1} have recently shown how modern 
functional renormalization group methods can be used 
to treat the full tangential momentum dependence of effective 
interactions for arbitrary curved Fermi surfaces.
Most recently, Salmhofer \cite{Sal1} has derived an improved
version of this field theoretic approach. The resulting flow
equations are particularly suitable for a concrete numerical
evaluation. 
To compute physical instabilities, we have derived the 
corresponding flow equations for susceptibilities \cite{HM1}.

In this letter we present results for the flow of susceptibilities 
as obtained by applying Salmhofer's renormalization group method 
to the two-dimensional Hubbard model with nearest and next-nearest
neighbor hopping on a square lattice.
The expected existence of a $d$-wave pairing instability driven by
antiferromagnetic spin fluctuations is thereby confirmed beyond 
doubt.
More unexpectedly, we find that strong forward scattering interactions
develop which may lead to a Pomeranchuk\cite{Pom} instability
breaking the tetragonal symmetry of the Fermi surface.


The one-band Hubbard model \cite{Mon}
\begin{equation}\label{HM}
 H = \sum_{\bi,\bj} \sum_{\sg} t_{\bi\bj} \,
 c^{\dag}_{\bi\sg} c_{\bj\sg} +
 U \sum_{\bj} n_{\bj\up} n_{\bj\down} \; ,
\end{equation}
describes tight-binding electrons with a local repulsion $U>0$.
Here $c^{\dag}_{\bi\sg}$ and $c_{\bi\sg}$ are creation and 
annihilation operators for fermions with spin projection 
$\sg \in \{ \up,\down \}$ on a lattice site ${\bi}$,
and $n_{\bj\sg} = c^{\dag}_{\bj\sg} c_{\bj\sg}$.
A hopping amplitude $-t$ between nearest neighbors and
an amplitude $-t'$ between next-nearest neighbors on a square
lattice leads to the dispersion relation
\begin{equation}
 \eps_{\bk} = -2t(\cos k_x + \cos k_y)
              -4t'\,\cos k_x \, \cos k_y
\end{equation}
for single-particle states.
This dispersion relation has saddle points at $\bk = (0,\pi)$
and $(\pi,0)$, which generate logarithmic van Hove 
singularities in the non-interacting density of states at the 
energy $\eps_{vH} = 4t'$.
For $t'=0$, $\eps_{\bk}$ has the nesting property 
$\eps_{\bk+\bQ} = - \eps_{\bk}$ for $\bQ = (\pi,\pi)$, which leads 
to an antiferromagnetic instability for arbitrarily small $U>0$ at
half-filling \cite{Mon}.


The RG equations are obtained as follows
(for details, see Salmhofer \cite{Sal1} and Ref.\ \cite{HM1}).
The infrared singularities are regularized by introducing an
infrared cutoff $\Lam > 0$ into the bare propagator such that
contributions from momenta with $|\eps_{\bk} - \mu| < \Lam$ 
are suppressed.
All Green functions of the interacting system will then flow as a 
function of $\Lam$, and the true theory is recovered in the limit 
$\Lam \to 0$.
Salmhofer \cite{Sal1} has recently pointed out that (amputated)
Green functions obtained by expanding the effective action of the 
theory in powers of {\em normal ordered}\/ monomials of fermion
fields obey differential flow equations with a structure that is 
particularly convenient for a power counting analysis to arbitrary 
loop order. 
With the bare interaction as initial condition at the highest 
scale $\Lam_0 = \max|\eps_{\bk}-\mu|$,
these flow equations determine the exact flow of the
effective interactions as $\Lam$ sweeps over the entire Brillouin
zone down to the Fermi surface.
The effective low-energy theory can thus be computed directly
from the microscopic model without introducing any ad hoc
parameters.

For a weak coupling stability analysis it is sufficient to 
truncate the exact hierarchy of flow equations at 1-loop 
level.
The effective 2-particle interaction then reduces to the 
one-particle irreducible 2-particle vertex $\Gam^{\Lam}$, and its 
flow is determined exclusively by $\Gam^{\Lam}$ itself (no higher 
many-particle interactions enter).
Flow equations for susceptibilities are obtained by considering
the exact RG equations in the presence of suitable external 
fields, which leads to an additional 1-particle term in the bare
interaction, and expanding everything in powers of the external
fields to sufficiently high order \cite{HM1}.

One cannot solve the flow equations with the full energy and momentum 
dependence of the vertex function, since $\Gam^{\Lam}$ has three 
independent energy and momentum variables.
The problem can however be much simplified by ignoring dependences 
which are {\em irrelevant} in the low energy limit, 
namely the energy dependence and the momentum dependence normal to 
the Fermi surface (for details, see Ref.\ \cite{HM1}).
This approximation is exact for the bare Hubbard vertex, and 
asymptotically exact in the low-energy regime.
The remaining tangential momentum dependence is discretized
for a numerical evaluation.
Most of our results where obtained for a discretization with 16 
points on the Fermi surface (yielding 880 ''running couplings''),
and we have checked that increasing the number of points does not 
change our results too much.


We have computed the flow of the vertex function for many 
different model parameters $t'$ and $U$ ($t$ just fixes the
absolute energy scale) and densities close to half-filling.
In all cases the vertex function develops a strong momentum
dependence for small $\Lam$ with divergencies for several
momenta at some critical scale $\Lam_c > 0$, which vanishes
exponentially for $U \to 0$.
To see which physical instability is associated with the
diverging vertex function we have computed commensurate and
incommensurate spin susceptibilities $\chi_S(\bq)$ with
$\bq = (\pi,\pi)$, $\bq = (\pi-\delta,\pi)$ and 
$\bq = (1-\delta) (\pi,\pi)$, where $\delta$ is a function 
of density \cite{Sch90}, the commensurate charge susceptibility 
$\chi_C(\pi,\pi)$, and singlet 
\begin{figure}
\centering\leavevmode\epsfbox{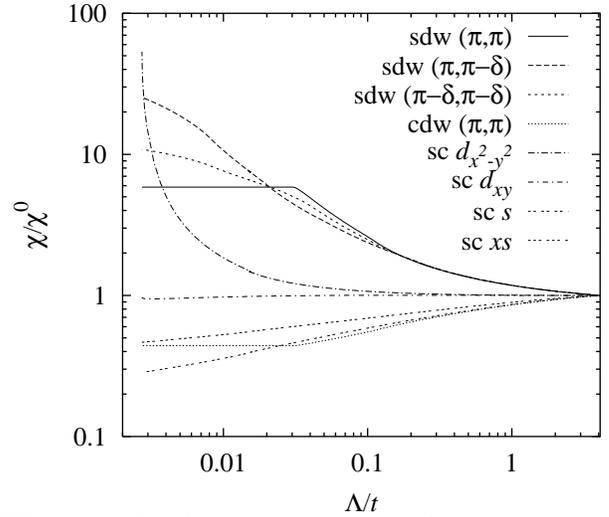}
\caption{The flow of the ratio of interacting and non-interacting 
 susceptibilities for $t' = -0.01t$, $U = t$ and $\mu = - 0.055t$.}
\label{fig1}
\end{figure}
\noindent
pair susceptibilities with form 
factors \cite{Sca}
\begin{equation}
 d(\bk) =  \left\{ \begin{array}{ll}
 1 & 
 \mbox{($s$-wave)} \\     
 \frac{1}{\sqrt{2}} (\cos k_x + \cos k_y) &
 \mbox{(extended $s$-wave)} \\
 \frac{1}{\sqrt{2}} (\cos k_x - \cos k_y) &
 \mbox{($d$-wave $d_{x^2-y^2}$)} \\
 \sin k_x \sin k_y &
 \mbox{($d$-wave $d_{xy}$)}. \end{array} \right.
\end{equation}
Some of these susceptibilities diverge together with the vertex
function at the scale $\Lam_c$.
Depending on the choice of $U$, $t'$ and $\mu$, the strongest
divergence is found for the commensurate or incommensurate
spin susceptibility or for the pair susceptibility with 
$d_{x^2-y^2}$ symmetry. In Fig.\ 1 we show a typical result for 
the flow of susceptibilities as a function of $\Lam$.
Note the threshold at $\Lam\approx 0.03t$ below which the amplitudes for 
various scattering processes, especially umklapp scattering, 
renormalize only very slowly. The flow of the antiferromagnetic 
spin susceptibility is cut off at the same scale. 
The {\em pairing susceptibility}\/ with $d_{x^2-y^2}$-symmetry is
obviously {\em dominant}\/ here (note the logarithmic scale).
Following the flow of the susceptibilities one can see that the 
$d_{x^2-y^2}$-pairing correlations develop in the presence of 
pronounced but {\em short-range antiferromagnetic spin-correlations}, 
in agreement with earlier ideas on $d$-wave superconductivity in
the Hubbard model \cite{Sca}.

In Fig.\ 2 we show the $(\mu,U)$ phase diagram for  $t' = - 0.01t$
obtained by identifying the dominant instability from the flow
for many different values of $\mu$ and $U$. 
For $\mu = 4t'$ the Fermi surface touches the saddle points 
$(0,\pi)$ and $(\pi,0)$, while $\mu = 4t'+0.01t$ corresponds to 
half-filling. 
Note that for $U \to 0$ the pairing instability always dominates, 
because the BCS channel dominates the flow in the limit 
$\Lam \to 0$. A spin density wave is the leading instability
for $U \to 0$ only in the special case with perfect nesting, $t'=0$ 
and $\mu=0$ (see the $(\mu,U)$ phase diagram computed from the 1-loop
flow for $t'=0$ in Ref.\ \cite{HM1}).
\begin{figure}
\centering\leavevmode\epsfbox{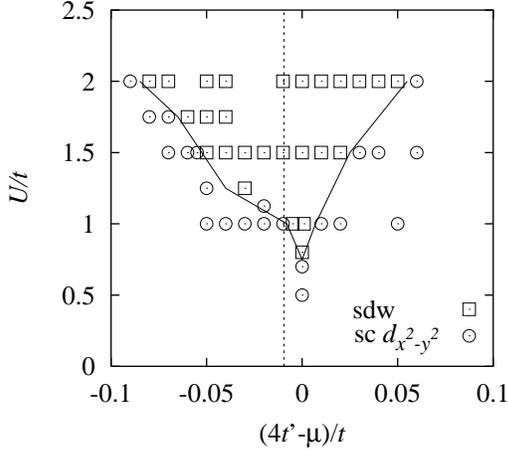}\vspace{1mm}
\caption{The $(\mu,U)$ phase diagram for $t'= -0.01t$ near 
half-filling (marked by the dashed vertical line); 
the symbols represent the parameter values for which the flow 
has been computed and whether the dominant instability is magnetic
(squares) or superconducting (circles); the solid line separates the
spin-density wave  regime from the superconducting regime.}
\label{fig2}
\end{figure}
\begin{figure}
\centering\leavevmode\epsfbox{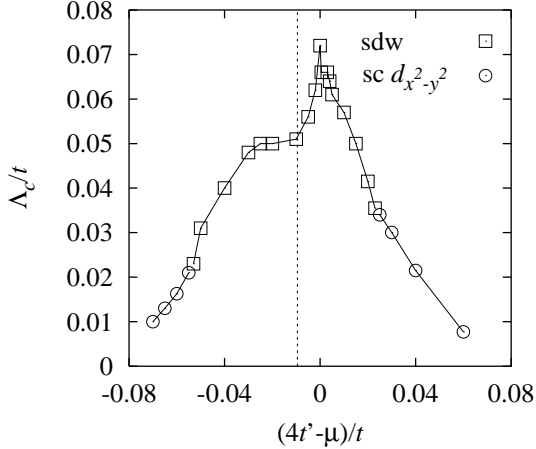}\vspace{1mm}
\caption{The critical energy scale $\Lam_c$ as a function of
 the chemical potential $\mu$ for $U = 1.5t$ and $t'= -0.01t$. 
 The different symbols indicate whether the leading instability is a
 spin-density wave or $d$-wave pairing instability.} 
\label{fig3}
\end{figure}
How the critical energy scale $\Lam_c$ varies as a function of
the chemical potential (i.e.\ as a function of density) is shown 
in Fig.\ 3 for an interaction strength $U = 1.5t$.
Obviously $\Lam_c$ is maximal for a chemical potential at the van
Hove energy.
Note that $\Lam_c$ must not be interpreted as a transition 
temperature for spin density wave formation or superconductivity,
but rather as an energy scale where bound particle-hole or
particle-particle pairs are formed.


Since some of the forward scattering interactions grow strong for
small $\Lam$, while the Fermi velocity is very small near the saddle
points, the Fermi surface may be significantly deformed by 
interactions, especially for $\mu \approx \eps_{vH}$.
\begin{figure}
\epsfxsize8cm
\centering\leavevmode\epsfbox{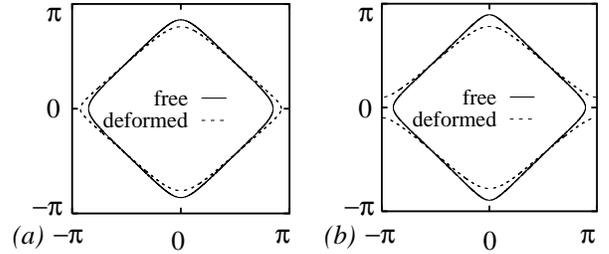}
\caption{Schematic plot of Fermi surface deformations breaking
 the square symmetry; the deformed surface may be closed (a) or
 open (b).}
\label{fig4}
\end{figure}
Previous investigations of Fermi surface deformations within
standard perturbation theory have yielded only very small shifts
even for sizable interaction strengths \cite{HM2}, but in these
studies the possibility of a spontaneous breaking of the point
group symmetry of the square lattice has not been taken into
account.

To analyze systematically the stability of the Fermi surface shape, 
we define a susceptibility
\begin{equation}
 \kappa_{\bk_F \bk'_F} = 
 \frac{\delta s_{\bk_F}}{\delta\mu_{\bk'_F}}
\end{equation}
which measures the size of Fermi surface shifts $\delta s_{\bk_F}$
for small momentum dependent shifts of the chemical potential 
$\delta\mu_{\bk'_F}$ at points $\bk'_F$ on the Fermi surface.
The matrix $\kappa_{\bk_F \bk'_F}$ defines a linear integral
operator acting on functions of $\bk_F$.
A simple consideration in the spirit of phenomenological Fermi
liquid theory shows that the corresponding inverse operator is 
given by
\begin{equation}
 (\kappa^{-1})_{\bk_F \bk'_F} =
 v_{\bk_F} \delta(\bk_F-\bk'_F) + 2 f^c_{\bk_F\bk'_F}
\end{equation}
where $v_{\bk_F}$ is the Fermi velocity and $f^c_{\bk_F\bk'_F}$
is the Landau function in the charge (spin-symmetric) channel.
It is now obvious that the matrix $\kappa_{\bk_F \bk'_F}$ is
symmetric. 
The Fermi surface is stable, if all eigenvalues of $\kappa$ (or
$\kappa^{-1}$) are positive. 
Note that Landau's energy functional can be written as 
a quadratic form in $\delta s_{\bk_F}$, with $\kappa^{-1}$ as 
kernel \cite{Noz}, and negative eigenvalues would imply that
this energy can be lowered by a suitable deformation of the
Fermi surface.
In isotropic Fermi liquids such instabilities occur for strongly 
negative Landau parameters, as first pointed out by Pomeranchuk 
\cite{Pom}.  

We have computed the renormalization group flow of the eigenvalues
and eigenvectors of the operator $\kappa^{-1}$ from the flow of 
the Landau function $f^{c\Lam}_{\bk_F\bk'_F}$, which is given 
directly by the vertex function in the forward scattering channel 
\cite{MCD}. 
For various choices of the model parameters we have always found 
a Fermi surface instability at a scale $\Lam_c^P$ above the scale 
$\Lam_c$ where the vertex function diverges.
In all cases the instability favors a deformation of the Fermi 
surface which breaks the point group symmetry of the square lattice, 
as shown schematically in Fig.\ 4.
The instability is mainly driven by a strong attractive
interaction between particles (or holes) on opposite corners of 
the Fermi surface near the saddle points and a repulsive interaction 
between particles on neighboring corners.

The above diagnosis of Pomeranchuk instabilities would be rigorous 
for a normal Fermi liquid with finite renormalized interactions in 
the infrared limit.
In the present system, however, the vertex function diverges at
a finite scale and possible Pomeranchuk type instabilities compete 
with magnetic and superconducting instabilities.
Since we have no quantitative theory of the strong coupling physics
near and below the scale $\Lam_c$, we can only list and discuss
two possible scenarios: \\
i) Energy gaps due to particle-particle or particle-hole binding
may stop the flow of forward scattering interactions before a 
Pomeranchuk instability sets in. \\
ii) The Pomeranchuk instability is not blocked by binding 
phenomena. 
In that case one would have a finite temperature phase transition 
with a spontaneous breaking of the (discrete) tetragonal symmetry 
of the square lattice, and subsequent continuous symmetry breaking
associated with magnetic order or superconductivity in the ground
state.

Which of the two scenarios is realized depends on the choice
of the model parameters.
The Pomeranchuk instability occurs more easily if the Fermi 
surface is close to the saddle points of $\eps_{\bk}$.
On the other hand, nesting raises the scale for particle-hole
binding (leading ultimately to magnetic order).
The best candidate is therefore the Hubbard model with a sizable 
$t'$ (reducing nesting) and $\mu = \eps_{vH}$.

We emphasize that the Pomeranchuk instability does not cut
off the singularity in the Cooper channel since it does not break
the reflection invariance. Hence, at sufficiently large doping
away from half-filling, $d$-wave superconductivity will set in in
any case, with an order parameter that may be slightly distorted
away from perfect $d$-wave symmetry.  
The Pomeranchuk instability would also not destroy the umklapp
scattering route to an insulating spin liquid discussed recently 
by Furukawa et al.\ \cite{FRS}.

To our knowledge a Pomeranchuk instability has not yet been
observed in numerical solutions of the two-dimensional Hubbard 
model. Of course this may be due to finite size limitations or 
too high temperatures in Monte Carlo simulations.
It would thus be interesting to compute the Fermi surface
susceptibility $\kappa_{\bk_F \bk'_F}$ numerically.

In real systems a Pomeranchuk instability as in Fig.\ 4 may 
lead to an orthorhombic lattice distortion, as a consequence of the 
coupling of electronic and lattice degrees of freedom.
High temperature superconductors indeed exhibit structural phase
transitions between tetragonal and orthorhombic phases.
It would be interesting to clarify whether a Pomeranchuk
instability might drive (at least to a significant extent) the 
transition into the orthorhombic phase in these materials. 


In summary, we have shown that modern renormalization group 
methods can be used to establish the expected $d$-wave pairing 
instability in the two-dimensional Hubbard beyond doubt.
Note that for small bare interactions and in a parameter
regime where only particle-particle pairing fluctuations grow
strong, the strong coupling problem associated with the formation 
of a superconducting state can be treated rigorously \cite{FMRT}.
Furthermore, we have pointed out that a Pomeranchuk instability
breaking the tetragonal symmetry of the Fermi surface is likely
to occur for a suitable choice of the model parameters, 
especially for a Fermi surface close to the saddle points in
the absence of perfect nesting.

\vspace{4mm}
\noindent{\bf Acknowledgments:} \\
We are very grateful to Maurice Rice, Manfred Salmhofer and 
Eugene Trubo\-witz for valuable discussions. 
This work has been supported by the Deutsche Forschungsgemeinschaft 
under Contract No.\ Me 1255/4 and within the Sonderforschungsbereich
341.


\widetext

\end{document}